https://doi.org/10.46813/2021-133-098# CROSS-SECTIONS OF PHOTONUCLEAR REACTIONS ON $^{nat}$Mo TARGETS AT END-POINT BREMSSTRAHLUNG ENERGY UP TO $E_{\gamma max}$ = 100 MeV

A.N. Vodin, O.S. Deiev, I.S. Timchenko, S.M. Olejnik, V.A. Kushnir, V.V. Mytrochenko,
S.O. Perezhogin, V.O. Bocharov
*National Science Center "Kharkov Institute of Physics and Technology", Kharkiv, Ukraine*
*E-mail: timchenko@kipt.kharkov.ua*Experiments to determine the yields and bremsstrahlung flux-averaged cross-sections $\langle\sigma(E_{\gamma max})\rangle$ of photonuclear reactions on the natural Mo targets were performed on the beam from the electron linear accelerator LUE-40 with the use of the γ-activation technique. The bremsstrahlung end-point energies were in the range $E_{\gamma max}$ = 35…80 MeV. The bremsstrahlung quantum flux was calculated with the program GEANT4.9.2 and, in addition, was monitored using the $^{100}$Mo(γ, n)$^{99}$Mo reaction. Calculations of the yields and average cross-sections $\langle\sigma(E_{\gamma max})\rangle$ for photonuclear reactions on stable Mo isotopes were computed using the σ(E) cross-sections from the TALYS1.95 code (for the level density model *LD*1). A comparison of experimental and calculated cross-sections $\langle\sigma(E_{\gamma max})\rangle$ for reactions $^{92}$Mo(γ, 2n)$^{90}$Mo and $^{92}$Mo(γ, pn)$^{90}$Nb was performed.
PACS: 25.20.-x, 27.60.+j## INTRODUCTION

At present, experimental studies of photodisintegration of nuclei in the photon energy range above the GDR and up to the threshold of pion production ($E_{th}$ = 145 MeV) are being actively carried out [1 - 7]. The interest for this energy range is due to the change in the mechanism of interaction of photons with nuclei: photodisintegration of nuclei through excitation of GDR and quasi-deuteron photoabsorption. However, the general shortage of experimental data in this energy range severely restricts both the general insight into the processes of γ-quantum interaction with nuclei and the model-approach testing capabilities.

Photodisintegration of molybdenum isotopes in the GDR region was investigated in early works [8, 9]. Investigations in these works were carried out both on bremsstrahlung gamma-ray with registration of the induced activity of the irradiated sample [8], and on quasi-monoenergetic photon beams with direct registration of photoneutrons [9]. However, in this method, the detected neutron cannot be unambiguously assigned to any of the reactions (γ, n), (γ, np) or (γ, n2p). A similar situation takes place when registering a proton in the reactions (γ, p), (γ, np), (γ, 2np). This leads to an ambiguous interpretation of the results and discrepancies in data from different laboratories.

In works [1, 2], the values of the relative yields for multiparticle reactions on natural molybdenum for bremsstrahlung energy of 67.7 MeV were determined and a comparison with theory was performed. The main difficulties in working with natural molybdenum targets, which associated with the presence of several stable isotopes of Mo with A = 92, 94-98, and 100, which lead to the production of the same nucleus, are also described.

The present work is concerned with the measurements of yields and bremsstrahlung flux-average cross-sections $\langle\sigma(E_{\gamma max})\rangle$ of photonuclear reactions on the natural Mo targets in the bremsstrahlung end-point energy range $E_{\gamma max}$ = 35…80 MeV. The comparisons of experimental and calculated yields and cross-sections $\langle\sigma(E_{\gamma max})\rangle$ for reactions $^{92}$Mo(γ, 2n)$^{90}$Mo and $^{92}$Mo(γ, pn)$^{90}$Nb were performed.

## 1. EXPERIMENTAL PROCEDURE

The experiments were performed using the bremsstrahlung gamma-beam from the LUE-40 RDC "Accelerator" NSC KIPT electron linear accelerator using the method of induced activity of the final product nucleus of the reaction. The experimental procedure is described in detail in [3, 4, 10].

The experimental complex for investigating photonuclear reactions is presented in the form of a block diagram in Fig. 1.

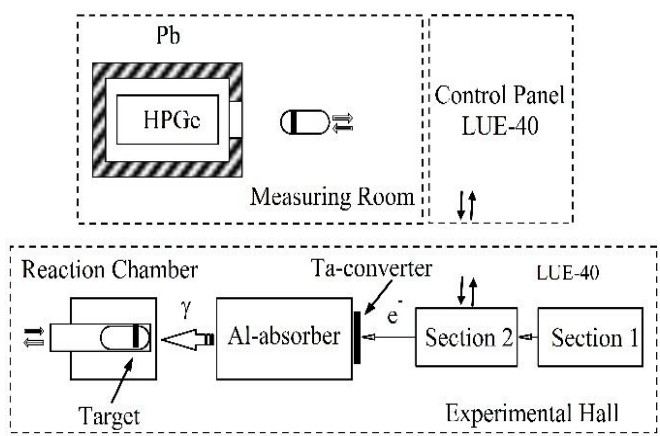

*Fig. 1. Experimental schematic diagram including three units shown with a dashed line. Above – the measuring room and the control panel of the LUE-40 accelerator, below – the experimental hall*

The studies for 10 values of electron energies were carried out at $E_e$ = 35.1, 39.9, 45.3, 50.0, 55.2, 60.1, 64.6, 70.3, 75.0, 80.7 MeV. The average beam current $I_e \approx 3$ μA. The electron energy spectrum width at FWHM makes $\Delta E_e/E_e \sim 1\%$. The bremsstrahlung gamma radiation was generated by passing a pulsed electron beam through a tantalum metal plate, 1.05 mm

98     *ISSN 1562-6016. BAHT. 2021. №3(133)*

in thickness. The Ta converter was fixed on the aluminum cylinder, 100 mm in diameter and 150 mm in thickness.

For the experiments, the natural molybdenum targets, which represented thin discs with diameters 8 mm and the thicknesses of 0.1 mm, were prepared. The target masses were $m \approx 60$ mg. To transport the capsule with the sample between the measuring room and the place of irradiation a pneumatic transport system was used.

The γ-quanta of the reaction products were detected using a Canberra GC-2018 semiconductor HPGe detector with the relative detection efficiency of 20%. The resolution FWHM is 1.8 keV for energy $E_\gamma = 1332$ keV and is 0.8 keV for $E_\gamma = 122$ keV. The dead time for γ-quanta detection varied between 0.1…5%. The absolute detection efficiency for γ-quanta of different energies was obtained using a standard set of γ-quanta sources: $^{241}$Am, $^{133}$Ba, $^{60}$Co, $^{137}$Cs, $^{22}$Na, $^{152}$Eu.

The bremsstrahlung flux was monitored by the yield of the $^{100}$Mo(γ,n)$^{99}$Mo reaction. For this purpose, was used γ-lines with an energy of $E_\gamma = 739.5$ keV, $T_{1/2} =$ 65.94 h, $I_\gamma = 12.13\%$ [11]. This approach made it possible to estimate the deviation of the real flux of bremsstrahlung from the calculated one [3, 12].

## 2. NATURAL MOLYBDENUM RADIATION SPECTRA ANALYSIS

The γ-radiation spectrum of a natural molybdenum target irradiated with high-energy γ-quanta is a complex pattern of emission lines of the $^{nat}$Mo(γ, xnyp) reactions located on a background substrate, which is formed as a result of Compton scattering of photons. As an example, Fig. 2 shows the spectrum of γ-radiation of a target with a mass of 57.7 mg after irradiation with the end-point bremsstrahlung energy $E_{\gamma max} = 60.1$ MeV.

The present work is concerned with the measurements of yields and average cross-sections $\langle\sigma(E_{\gamma max})\rangle$ of photonuclear reactions $^{92}$Mo(γ, 2n)$^{90}$Mo and $^{92}$Mo(γ, pn)$^{90}$Nb on the natural Mo targets in the energy range $E_{\gamma max} = 35…80$ MeV. The characteristics of the reactions are presented in Table 1 according to [11].

*Table 1*

*Nuclear spectroscopic data of the radio-nuclides reactions from [11]*

| Reaction | $T_{1/2}$, h | $E_\gamma$, keV ($I_\gamma$, %) |
|---|---|---|
| $^{92}$Mo(γ, 2n)$^{90}$Mo | 5.56±0.09 | 122.37 (64.2) 257.34 (78) |
| $^{92}$Mo(γ, pn)$^{90}$Nb | 14.60±0.05 | 1129.224 (92.7) |
| $^{100}$Mo(γ, n)$^{99}$Mo | 65.94±0.01 | 739.50 (12.13) |

The self-absorption coefficient in the target for the 122.37 keV line does not exceed 4.4%, and for 257.34 keV − 1%. This coefficient was taken into account when processing the results of the experiment.

Natural molybdenum consists of 7 stable isotopes, the isotope percent abundance of which was taken from the database [13, 14]: 92 – 14.84%, 94 – 9.25%, 95 – 15.92%, 96 – 16.68%, 97 – 9.55%, 98 – 24.13%, 100 – 9.63%. This is somewhat different from the values used in the works [1, 2].

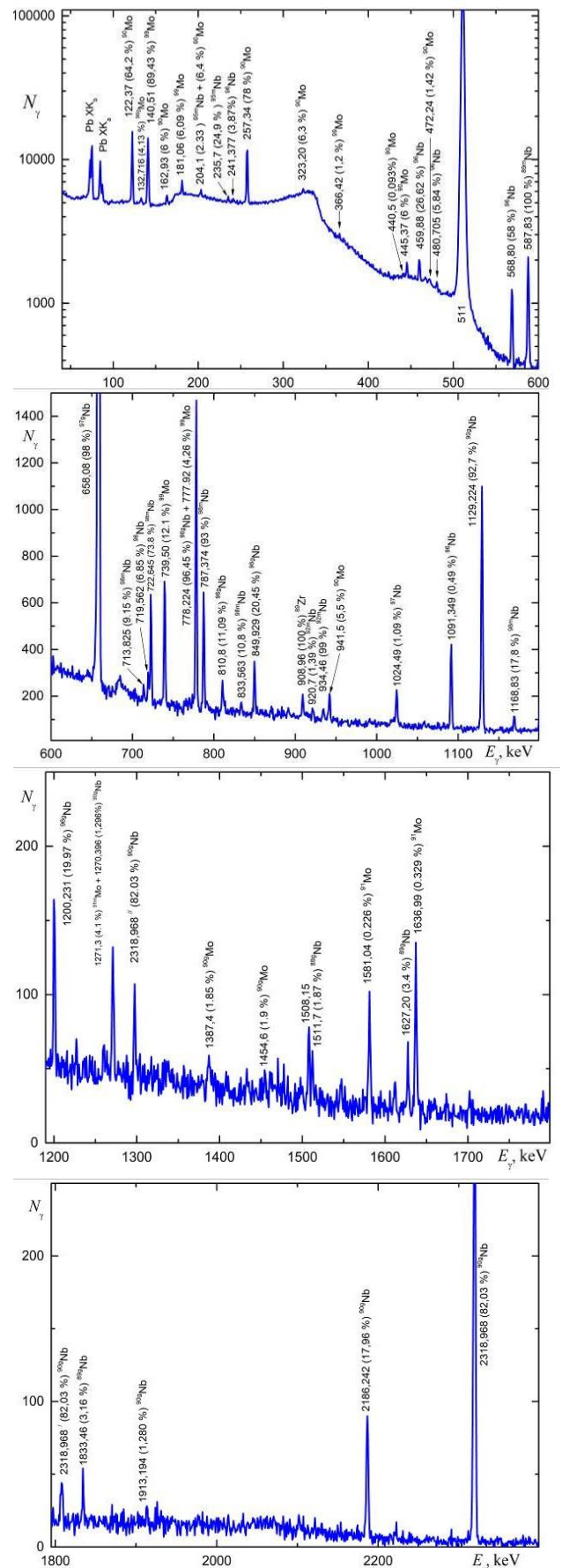

*Fig. 2. Spectrum of γ-radiation of a $^{nat}$Mo target with a mass of 57.7 mg after irradiation with the end-point bremsstrahlung energy $E_{\gamma max} = 60.1$ MeV. The irradiation and measurement times are 30 min*

It should be noted that $^{90}$Mo and $^{90}$Nb nuclei could have been formed in 7 different reactions on isotopes of $^{nat}$Mo. The thresholds of these reactions varied in the energy range $E_{th}$ = 22.8…86.8 MeV for the case of $^{90}$Mo production, and $E_{th}$ = 17.3…83.5 MeV for the case of $^{90}$Nb production.

## 3. CALCULATIONS OF CROSS-SECTIONS FOR PHOTONUCLEAR REACTIONS USING TALYS1.95 (*LD*1) AND GEANT4.9.2 CODES

The electron bremsstrahlung spectra were calculated using the open-source software code GEANT4.9.2, PhysList G4LowEnergy [14]. The real geometry of the experiment was used in calculations also the space and energy distributions of the electron beam were taken into account. Fig. 3 shows the calculated bremsstrahlung spectra, which were used in calculations of the gamma-flux irradiating the target.

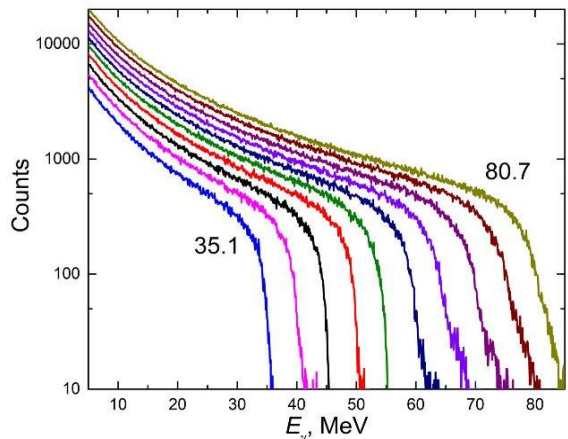

*Fig. 3. Bremsstrahlung spectra calculated in GEANT4.9.2 for electron energies $E_e$=35.1, 39.9, 45.3, 50.0, 55.2, 60.1, 64.6, 70.3, 75.0, 80.7 MeV*

The calculation of the cross-sections $\sigma(E)$ for the reactions $^{nat}$Mo($\gamma$,$xn$)$^{90}$Mo and $^{nat}$Mo($\gamma$,$pxn$)$^{90}$Nb for monochromatic photons was performed using the TALYS1.95 code [13], which was installed on Linux Ubuntu-20.04. The calculations were performed for the *LD*1 level density model: Constant temperature + Fermi gas model (Figs. 4,a and 5,a).

The calculated cross-sections $\sigma(E)$ were then averaged over the bremsstrahlung flux $W(E, E_{\gamma max})$ in the energy range from the threshold $E_{th}$ of the corresponding reaction channel to the maximum energy of the bremsstrahlung $\gamma$-quanta spectrum $E_{\gamma max}$ = 35…80 MeV. As a result, the values of the bremsstrahlung flux-average cross-sections were obtained:

$$\langle \sigma(E_{\gamma\max}) \rangle = \frac{\int_{E_{th}}^{E_{\gamma\max}} \sigma(E) W(E, E_{\gamma\max}) dE}{\int_{E_{th}}^{E_{\gamma\max}} W(E, E_{\gamma\max}) dE}. \quad (1)$$

The calculated in this way $\langle \sigma(E_{\gamma max}) \rangle$ values were compared with the experimental measured average cross-sections determined by the expression:

$$\langle \sigma(E_{\gamma\max}) \rangle = \frac{\lambda \Delta A}{\varepsilon N_x I_\gamma \Phi(E_{\gamma\max})(1-e^{-\lambda t_{irr}})e^{-\lambda t_{cool}}(1-e^{-\lambda t_{meas}})}, \quad (2)$$

where $\Delta A$ is the number of counts of $\gamma$-quanta in the full absorption peak (for the $\gamma$-line of the investigated reaction),

$$\Phi(E_{\gamma\max}) = \int_{E_{th}}^{E_{\gamma\max}} W(E, E_{\gamma\max}) dE$$

is the bremsstrahlung quanta flux in the energy range from the reaction threshold $E_{th}$ up to $E_{\gamma max}$; $N_x$ is the number of investigated atoms; $I_\gamma$ – the absolute intensity of the analyzed $\gamma$-quanta; $\varepsilon$ – the absolute detection efficiency for the analyzed $\gamma$-quanta energy; $\lambda$ is the decay constant (ln2/$T_{1/2}$); $t_{irr}$, $t_{cool}$, and $t_{meas}$ are the irradiation time, cooling time and measurement time, respectively. From eq. (1) and (2) it follows that the value of the average cross-section $\langle \sigma(E_{\gamma max}) \rangle$ depends on the energy distribution of the bremsstrahlung flux and on the value of the reaction threshold $E_{th}$.

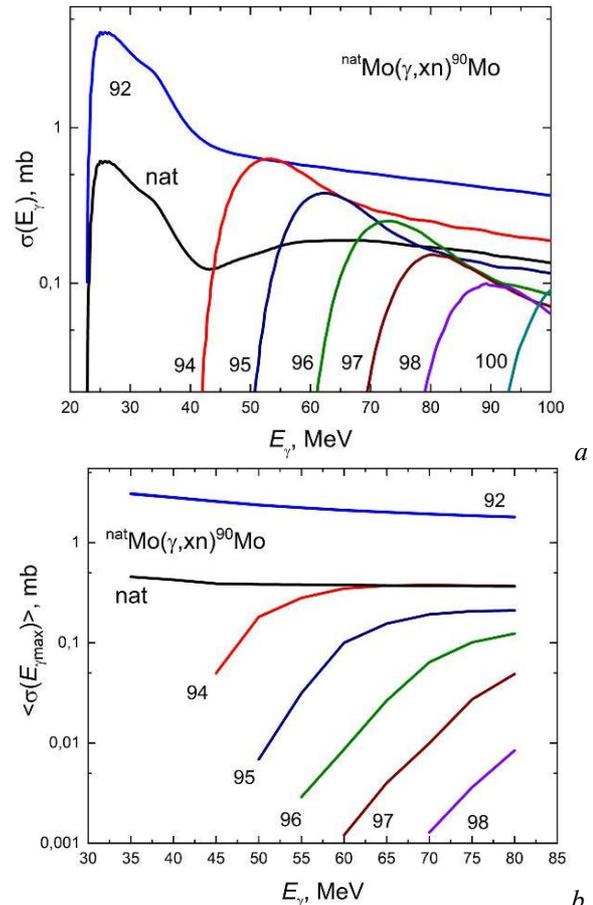

*Fig. 4. TALYS1.95 (LD1) computation of cross-sections $\sigma(E)$ production $^{90}$Mo for $^{nat}$Mo($\gamma$, xn)$^{90}$Mo reactions at different isotopes (92, 94-98, 100). The total cross-section (black curve) is calculated taking into account the percentage contribution of isotopes (a); bremsstrahlung flux-averaged cross-section $\langle \sigma(E_{\gamma max}) \rangle$ for $^{nat}$Mo($\gamma$,xn)$^{90}$Mo at different isotopes (92, 94-98). The total cross-section (black curve) is calculated taking into account the percentage contribution of isotopes (b)*

The calculations of $\langle \sigma(E_{\gamma max}) \rangle$ using real bremsstrahlung spectra for the reactions $^{nat}$Mo($\gamma$, $xn$)$^{90}$Mo and $^{nat}$Mo($\gamma$, $pxn$)$^{90}$Nb are shown in Figs. 4,b and 5,b, respectively.

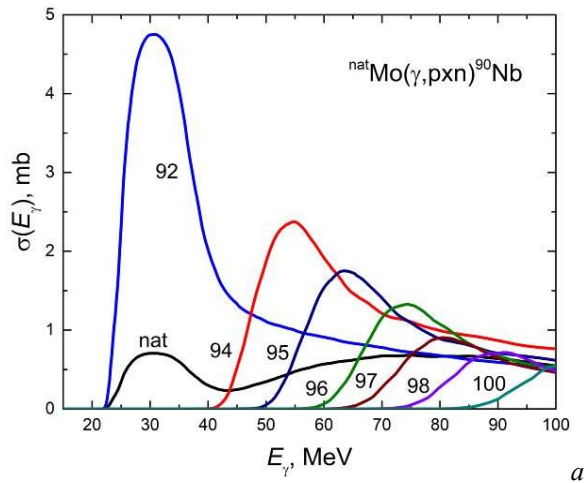

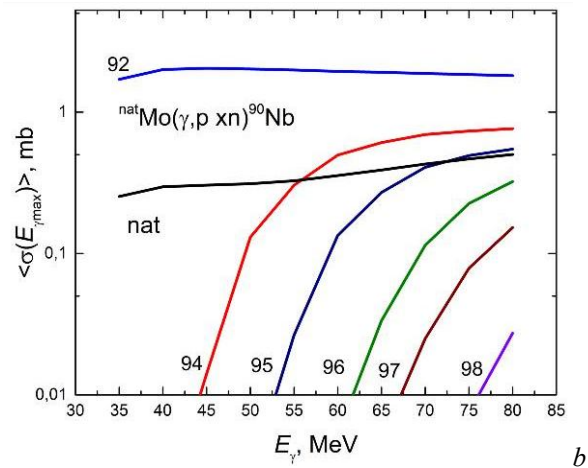

*Fig. 5. TALYS1.95 (LD1) computation of cross-sections σ(E) production $^{90}$Nb for $^{nat}$Mo(γ, pxn)$^{90}$Nb reactions at different isotopes (92, 94-98, 100) (a); bremsstrahlung flux-averaged cross-section ⟨σ(E$_{γmax}$)⟩ for $^{nat}$Mo(γ, pxn)$^{90}$Nb at different isotopes (92, 94-98). The total cross-section (black curve) is calculated taking into account the percentage contribution of isotopes (b)*

The reaction yield is defined as:

$$Y(E_{\gamma\max}) = N_x \int_{E_{th}}^{E_{\gamma\max}} \sigma(E) W(E, E_{\gamma\max}) dE. \quad (3)$$

This value is used in photonuclear experiments and is convenient for estimating the contributions of the reaction channels to the total reaction yield.

For the reactions $^{nat}$Mo(γ, xn)$^{90}$Mo, $^{nat}$Mo(γ, pxn)$^{90}$Nb, the dominant channels are $^{92}$Mo(γ, 2n)$^{90}$Mo and $^{92}$Mo(γ, pn)$^{90}$Nb, respectively. The contribution of the channels with higher $E_{th}$ can be estimated from Figs. 4 and 5. Estimates of the contributions (K) of dominant reactions to the yield of production of the $^{90}$Mo and $^{90}$Nb nuclei, calculated using TALYS1.95 (LD1) code, are presented in Table 2. These values are valid for the total average cross-sections calculated by averaging the total cross-section σ(E) with the minimum $E_{th}$, i.e., for $^{nat}$Mo(γ, xn)$^{90}$Mo and $^{nat}$Mo(γ, pxn)$^{90}$Nb, the values $E_{th}$ = 22.8 and 17.3 MeV, respectively.

The values of the coefficients K, given in Table 2 were used to estimate the experimental values of the yields and average cross-sections of the $^{92}$Mo(γ, 2n)$^{90}$Mo and $^{92}$Mo(γ, pn)$^{90}$Nb.

*Table 2*
*Contributions (K) of dominant reactions to the yields of production of the $^{90}$Mo and $^{90}$Nb nuclei in photonuclear reactions on natural molybdenum*

| $E_{γmax}$, MeV | $^{92}$Mo(γ, 2n)$^{90}$Mo | $^{92}$Mo(γ, pn)$^{90}$Nb |
|---|---|---|
| 35.1 | 1 | 1 |
| 39.9 | 1 | 1 |
| 45.3 | 1 | 1 |
| 50 | 0.992 | 0.984 |
| 55.2 | 0.981 | 0.957 |
| 60.1 | 0.965 | 0.913 |
| 64.6 | 0.949 | 0.871 |
| 70.3 | 0.932 | 0.825 |
| 75 | 0.917 | 0.785 |
| 80.7 | 0.905 | 0.751 |

## 4. RESULTS AND DISCUSSION

For the reactions $^{92}$Mo(γ, 2n)$^{90}$Mo and $^{92}$Mo(γ, pn)$^{90}$Nb, the yields $Y(E_{γmax})$ were experimentally determined in the energy range $E_{γmax}$ = 35…80 MeV (Fig. 6). The used γ-lines were 122.37, 257.34 keV for $^{92}$Mo(γ, 2n)$^{90}$Mo and 1129.224 keV for $^{92}$Mo(γ, pn)$^{90}$Nb (see Table 1). The experimental reaction yields were multiplied by the corresponding K factors (see Table 2). Comparison shows that the yield of the reaction with the production of the $^{90}$Nb nucleus is 1.5 times higher than in the case of the production of the $^{90}$Mo nucleus.

The yields of the studied reactions have also been calculated using the σ(E) from TALYS1.95 (LD1) code. Fig. 6 shows that there is a noticeable (1.5…2 times) excess of the experimental reaction yields over the calculated ones. In the case of a reaction with a charged particle in the exit channel, this difference is higher.

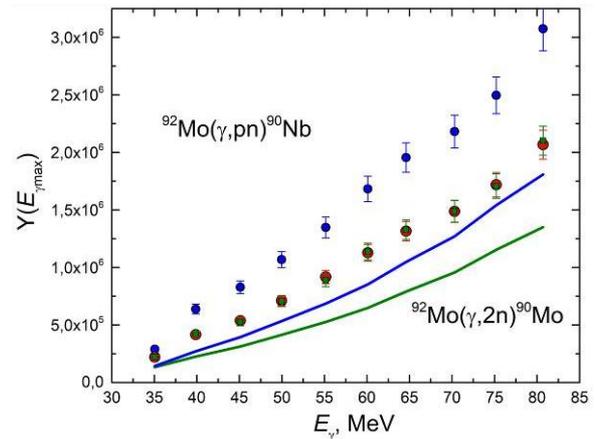

*Fig. 6. The yields Y(E$_{γmax}$) for the reactions $^{92}$Mo(γ, 2n)$^{90}$Mo and $^{92}$Mo(γ, pn)$^{90}$Nb: TALYS1.95 (LD1) computations – green and blue curves; experimental reaction $^{92}$Mo(γ, 2n)$^{90}$Mo yields: red circles – 257.34 keV, green squares – 122.37 keV; yield $^{92}$Mo(γ, pn)$^{90}$Nb: blue circles – 1129.224 keV*

The average cross sections for the reactions $^{92}$Mo(γ, 2n)$^{90}$Mo and $^{92}$Mo(γ, pn)$^{90}$Nb were experimentally determined and calculated in the TALYS1.95 (LD1) code. The results of comparing the experimental and calculated cross-sections ⟨σ(E$_{γmax}$)⟩ for both reactions are shown in Fig. 7,a,b. The analysis of the differ-ences between the experimental and calculated ⟨σ($E_{γmax}$)⟩ values above 55 MeV was carried out with a correction for the coefficients $K$ (which were calculated with the σ($E$) from the TALYS1.95 code).

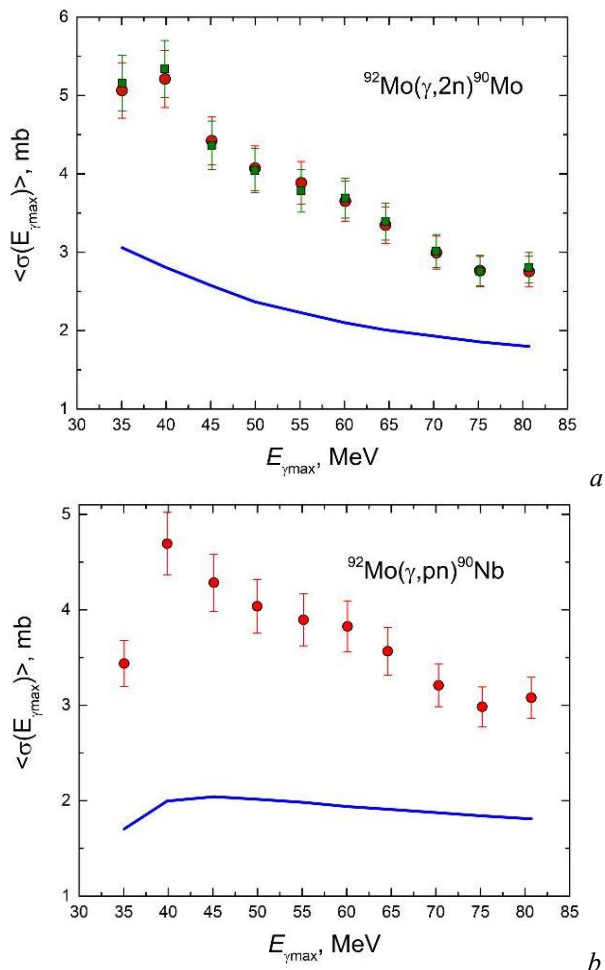

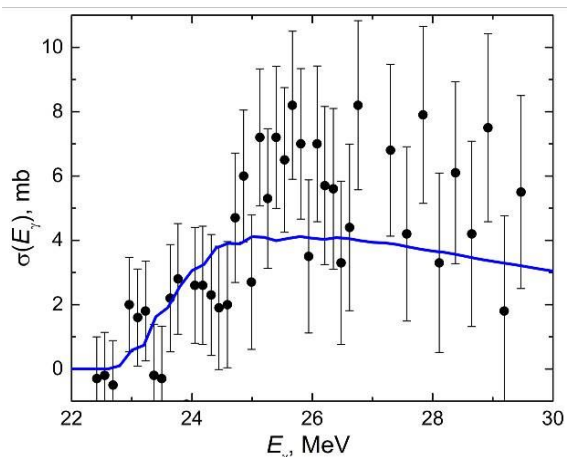

Fig. 8. Cross-section σ($E$) of the $^{92}Mo(γ, 2n)^{90}Mo$ reaction. Black circles – result from [9], blue curve – calculation in the code TALYS1.95 (LD1)

In the case of the $^{92}Mo(γ, pn)^{90}Nb$ reaction, the excess of the experimental ⟨σ($E_{γmax}$)⟩ over the calculated values is slightly greater: 2.1 times at 35 MeV and 1.7 times at 80 MeV. In this case, the calculation of the cross-section for the $^{92}Mo(γ, pn)^{90}Nb$ reaction in the TALYS1.95 (LD1) code is also underestimated.

## CONCLUSIONS

Experiments to determine the yields Y($E_{γmax}$) and bremsstrahlung flux-averaged cross-sections ⟨σ($E_{γmax}$)⟩ of photonuclear reactions for the natural Mo targets were performed on the beam from the electron linear accelerator LUE-40 with the use of the γ-activation technique. The bremsstrahlung end-point energies were in the range $E_{γmax}$ = 35…80 MeV.

For multiparticle reactions $^{nat}Mo(γ,xn)^{90}Mo$ and $^{nat}Mo(γ, pxn)^{90}Nb$, the cross-sections σ($E$) were calculated for Mo isotopes with A = 92, 94-98, and 100 in the range $E_{γmax}$ up to 100 MeV in the TALYS1.95 code for level density models LD1. These values are used to calculate the yields and average cross-sections.

It is shown that in the studied energy range in the reaction $^{nat}Mo(γ, xn)^{90}Mo$, the dominant channel is $^{92}Mo(γ, 2n)^{90}Mo$. The contribution of this reaction to the total value of the formation of the $^{90}Mo$ nucleus was 90% at 80 MeV. At the same time, for the case of the reaction $^{nat}Mo(γ, pxn)^{90}Nb$, the contribution of the dominant channel of the reaction $^{92}Mo(γ, pn)^{90}Nb$ decreases faster with increasing energy and at 80 MeV was 75%.

Comparison of the experimental and calculated values of ⟨σ($E_{γmax}$)⟩ for the reaction $^{92}Mo(γ, 2n)^{90}Mo$, showed a noticeable excess (up to two times) of the experimental results over the TALYS1.95 estimates. This can be explained by the underestimation of the cross-section σ($E$) from the TALYS1.95 (LD1) code. A similar result was obtained by comparing the experimental and calculated ⟨σ($E_{γmax}$)⟩ for the $^{92}Mo(γ, pn)^{90}Nb$ reaction.

Fig. 7. The cross-section ⟨σ($E_{γmax}$)⟩ for the reactions $^{92}Mo(γ, 2n)^{90}Mo$: TALYS1.95 (LD1) computation – blue curves; experimental value: red circles – 257.34 keV, green squares – 122.37 keV (a);
the cross-section ⟨σ($E_{γmax}$)⟩ for the reactions $^{92}Mo(γ, pn)^{90}Nb$: TALYS1.95 (LD1) computation – blue curves; experimental value: red circles – 1129.224 keV (b)

As it can be seen from these figures, the experimental values exceed the calculated values ⟨σ($E_{γmax}$)⟩. So, in the case of the reaction $^{92}Mo(γ, 2n)^{90}Mo$, the difference was 1.7 times at an energy of 35 MeV and decreases with increasing energy up to 1.5 times. The values of the total average cross-sections at the energy $E_{γmax}$ = 35…55 MeV, according to Table 2 are determined only by the contribution of the reaction $^{92}Mo(γ, 2n)^{90}Mo$. Consequently, the deviation of the experimental values from the calculated ones at these energies indicates an underestimation of the cross-section in the TALYS1.95 (LD1) code.

To verify this statement, Fig. 8 shows the data from [9] for the partial cross-section σ($E$) of the $^{92}Mo(γ, 2n)^{90}Mo$ reaction. In this case, the experimental values above 25 MeV are systematically higher than the calculation in the TALYS1.95.

## REFERENCES

1. B.S. Ishkhanov, I.M. Kapitonov, A.A. Kuznetsov, et al. Photodisintegration of molybdenum isotopes // *Moscow University Physics Bulletin*. 2014, № 1, p. 35-43.

2. B.S. Ishkhanov, A.A. Kuznetsov, V.N. Orlin, et al. Photonuclear reactions on molybdenum isotopes // *Phys. At. Nucl.* 2014, v. 77, p. 1427-1435.
3. A.N. Vodin, O.S. Deiev, V.Yu. Korda, I.S. Timchenko, S.N. Olejnik, N.I. Aizatsky, A.S. Kachan, L.P. Korda, E.L. Kuplennikov, V.A. Kushnir, V.V. Mitrochenko, S.A. Perezhogin. Photoneutron reactions on $^{93}$Nb at $E_{\gamma max}$ = 33…93 MeV // arXiv:2101.08614.
4. A.N. Vodin, O.S. Deiev, I.S. Timchenko, S.N. Olejnik. Cross-sections for the $^{27}$Al($\gamma$, x)$^{24}$Na multiparticle reaction at $E_{\gamma max}$ = 40…95 MeV // arXiv:2012.14475.
5. A.N. Vodin, O.S. Deiev, I.S. Timchenko, S.N. Olejnik, M.I. Ayzatskiy, V.A. Kushnir, V.V. Mitrochenko, S.A. Perezhogin. Photoneutron Reactions Photoneutron Reactions $^{181}$Ta($\gamma$, xn; x=1…8)$^{181-x}$Ta at $E_{\gamma max}$ = 80…95 MeV // arXiv:2103.09859.
6. M. Zaman, G. Kim, H. Naik, et al. Flux weighted average cross-sections of $^{nat}$Ni($\gamma$, x) reactions with the bremsstrahlung end-point energies of 55, 59, 61 and 65 MeV // *Nucl. Phys. A.* 2018, v. 978, p. 173-186.
7. H. Naik, G. Kim, M. Zaman, et al. Photo-neutron reaction cross-sections of $^{59}$Co in the bremsstrahlung end-point energies of 65 and 75 MeV // *Eur. Phys. J. A* (2019) 55: 217 DOI 10.1140/epja/i2019-12915-y.
8. N. Mutsuro, Y. Ohnuki, K. Sato, M. Kimura. Photoneutron Cross Sections for Ag$^{107}$, Mo$^{92}$ and Zr$^{90}$ // *J. of the Phys. Soc. of Japan.* 1959, v. 14, p. 1649.
9. H. Beil, R. Bergire, P. Carlos, A. Lepretre, et al. A study of the photoneutron contribution to the giant dipole resonance in doubly even Mo isotopes // *Nucl. Phys. A.* 1974, v. 227, p. 427.
10. A.N. Vodin, O.S. Deiev, I.S. Timchenko, S.N. Olejnik, A.S. Kachan, L.P. Korda, E.L. Kuplennikov, V.A. Kushnir, V.V. Mitrochenko, S.A. Perezhogin, N.N. Pilipenko, V.S. Trubnikov. Cross-Sections for Photonuclear Reactions $^{93}$Nb($\gamma$,n)$^{92m}$Nb and 93Nb($\gamma$,n)$^{92t}$Nb in the end-Point Bremsstrahlung Energies 36…91 MeV // *Problems of Atomic Science and Technology.* 2020, № 3, p. 148-153.
11. S.Y.F. Chu, L.P. Ekstrom, R.B. Firestone. The Lund/LBNL, Nuclear Data Search, Version 2.0, February 1999, WWW Table of Radioactive Isotopes, available from http://nucleardata.nuclear.lu.se/toi/.
12. A.N. Dovbnua, A.S. Deiev, V.A. Kushnir, et al. Experimental results on cross sections for $^{7}$Be photoproduction on $^{12}$C, $^{14}$N, and $^{16}$O nuclei in the energy range of 40…90 MeV // *Physics of Atomic Nuclei.* 2014, v. 77, p. 805-808.
13. A. Koning and D. Rochman // *Nucl. Data Sheets.* 2012, v. 113, p. 2841; TALYS – based evaluated nuclear data library. https://tendl.web.psi.ch/tendl2019/tendl2019.html.
14. S. Agostinelli et al. // *Methods Phys. A.* 2003, v. 506, p. 250; Electron and Positron Incident. http://GEANT4.9.2.web.cern.ch/GEANT4.9.2/.

## СЕЧЕНИЯ ФОТОЯДЕРНЫХ РЕАКЦИЙ НА МИШЕНЯХ ИЗ $^{nat}$Mo ПРИ ЭНЕРГИИ ТОРМОЗНЫХ КВАНТОВ ДО $E_{\gamma max}$ = 100 МэВ

*А.Н. Водин, А.С. Деев, И.С. Тимченко, С.Н. Олейник, В.А. Кушнир, В.В. Митроченко, С.А. Пережогин, В.А. Бочаров*

Эксперименты по определению выходов и сечений фотоядерных реакций на мишенях из натурального Мо выполнены на пучке линейного ускорителя электронов LUE-40 с использованием γ-активационной методики. Область граничных энергий тормозных γ-квантов составляла $E_{\gamma max}$ = 35…80 МэВ. Поток тормозных квантов рассчитывался в GEANT4.9.2 и дополнительно мониторировался по выходу реакции $^{100}$Mo(γ, n)$^{99}$Mo. Расчеты выходов и средних сечений $\langle\sigma(E_{\gamma max})\rangle$ для фотоядерных реакций на стабильных изотопах Мо проводились с использованием сечений σ(E) из кода TALYS1.95 (для модели плотности уровней *LD*1). Проведено сравнение экспериментальных и расчетных значений $\langle\sigma(E_{\gamma max})\rangle$ для реакций $^{92}$Mo(γ, 2n)$^{90}$Mo, $^{92}$Mo(γ, pn)$^{90}$Nb.

## ПЕРЕРІЗИ ФОТОЯДЕРНИХ РЕАКЦІЙ НА МІШЕНЯХ З $^{nat}$Mo ПРИ ЕНЕРГІЇ ГАЛЬМІВНИХ КВАНТІВ ДО $E_{\gamma max}$ = 100 МеВ

*О.М. Водін, О.С. Деєв, І.С. Тімченко, С.М. Олійник, В.А. Кушнір, В.В. Мітроченко, С.О. Пережогін, В.О. Бочаров*

Експерименти по визначенню виходів і перетинів фотоядерних реакцій на мішенях з натурального Мо виконані на пучку лінійного прискорювача електронів LUE-40 з використанням γ-активаційної методики. Область граничних енергій гальмівних γ-квантів становила $E_{\gamma max}$ = 35…80 МеВ. Потік гальмівних квантів розраховувався в GEANT4.9.2 і додатково моніторувався по виходу реакції $^{100}$Mo(γ, n)$^{99}$Mo. Розрахунки виходів і середніх перетинів $\langle\sigma(E_{\gamma max})\rangle$ для фотоядерних реакцій на стабільних ізотопах Мо проводилися з використанням перетинів σ(E) з коду TALYS1.95 (для моделі щільності рівнів *LD*1). Проведено порівняння експериментальних і розрахункових $\langle\sigma(E_{\gamma max})\rangle$ для реакцій $^{92}$Mo(γ, 2n)$^{90}$Mo, $^{92}$Mo(γ, pn)$^{90}$Nb.